\documentclass[usenatbib]{mn2e}
\usepackage{graphicx}
\title[Oxygen abundances in the most oxygen-rich spiral galaxies]
      {Oxygen abundances in the most oxygen-rich spiral galaxies}

\author[L.S. Pilyugin, T.X. Thuan and J.M. V\'{\i}lchez]
       {Leonid S. Pilyugin$^{1}$, Trinh X. Thuan$^{2}$ and Jos\'{e} M. V\'{\i}lchez$^{3}$\\
       $^{1}$Main Astronomical Observatory
             of National Academy of Sciences of Ukraine,
             27 Zabolotnogo str., 03680 Kiev, Ukraine
              \\
       $^{2}$Astronomy Department, University of Virginia, P.O.Box 3818, 
             University Station, Charlottesville, VA 22903
             \\
       $^{3}$Instituto de Astrof\'{\i}sica de Andaluc\'{\i}a,
             CSIC, Apdo, 3004, 18080 Granada, Spain
       \\}

\date{Received 2005 November 28; in original form 2005 July 26}

\begin{document}

\maketitle

\begin{abstract}
Oxygen abundances in the spiral galaxies expected to be richest in 
oxygen are estimated. The new abundance 
determinations are based on the recently discovered ff -- relation  
between auroral and nebular oxygen line fluxes in high-metallicity 
H\,{\sc ii} regions. We find that the maximum gas-phase oxygen abundance 
in the central regions of spiral galaxies is 12+log(O/H) $\sim$ 8.75. 
This value is significantly lower (by a factor $\ga$ 5) than the previously 
accepted value. The central oxygen abundance in the Milky Way is similar 
to that in other large spirals.
\end{abstract}

\begin{keywords}
galaxies: abundances -- ISM: abundances -- H\,{\sc ii} regions
\end{keywords}

\section{Introduction}

The oxygen abundance in spiral galaxies has been  the subject of much 
discussion for a long 
time.  Which spiral galaxy is the oxygen-richest one? And how high is the 
oxygen abundance in the oxygen-richest galaxies? 
Due to the presence of radial abundance gradients in the disks of spiral galaxies, 
the maximum oxygen abundance occurs at their centers.  
 The notation (O/H)$_{\rm C}$ = 12+log(O/H)$_{R=0}$ will be used 
thereafter to denote the central oxygen abundance. 

The  chemical composition of various samples of spiral galaxies has been discussed in a 
number of papers \citep[e.g.][]{vila92,zkh,garnettetal97,vanzeeetal98,garnett02}.  
According to those articles, the oxygen-richest galaxies are: 
NGC~5194 (M~51) with (O/H)$_{\rm C}$ = 9.54 \citep{vila92}, 
NGC~3351 with (O/H)$_{\rm C}$ = 9.41 \citep{zkh}, NGC~3184 with 
(O/H)$_{\rm C}$ = 9.50 \citep{vanzeeetal98}, NGC~6744 with (O/H)$_{\rm C}$ = 
9.64 \citep{garnettetal97}. 
Different versions of the one-dimensional empirical method, 
proposed first by \citet{pageletal79} a quarter of a century ago, have been 
used for oxygen abundance determinations in those papers.
 \citet{lcal,hcal,m101,vybor} has shown that the oxygen abundances in galaxies 
determined with the one-dimensional empirical calibrations 
are significantly overestimated at the high-metallicity end ( 12 + log O/H $>$ 8.25).  
This for two reasons. First, the then-existing  calibrating 
points at the high-metallicity end were very few and not reliable. 
Second, the physical conditions in H\,{\sc ii} regions
cannot be taken into account accurately in one-dimensional calibrations.
\citet{pilvilcon04} have used instead a two-dimensional parametric empirical calibration
to derive oxygen abundances for a sample of spiral galaxies which  
includes NGC~3184, NGC~3351, NGC~5194, and NGC~6744. They found generally lower  
oxygen abundances with  (O/H)$_{\rm C}$ $\sim$ 9.0.  

Here we consider anew the problem of the maximum oxygen abundance in 
spiral galaxies by attempting to derive more accurate oxygen abundances. 
These can be derived via the 
classical T$_{\rm e}$ -- method, T$_{\rm e}$ being the electron temperature of the  
H\,{\sc ii} region. Measurements of the auroral lines, such as 
[OIII]$\lambda$4363, are necessary to determine T$_{\rm e}$. 
Unfortunately, they are very faint and often drop below the  
detectability level in the spectra of high-metallicity H\,{\sc ii} regions. 
\citet{ff} has advocated that the faint auroral line flux can be 
computed from the fluxes in the strong nebular lines via the ff -- relation.
Then, using the obtained flux in the auroral line, accurate oxygen 
abundances can be derived using the classical T$_{\rm e}$ -- method. 
We will estimate (O/H)$_{\rm C}$ in the spiral galaxies reported 
to be the oxygen-richest ones, NGC~3184, 
NGC~3351, NGC~5194, and NGC~6744. For comparison, 
we will also consider  NGC~2903, 
NGC~2997, NGC~5236 as well the Milky Way Galaxy.

We describe the method used to determine oxygen abundances in  
H\,{\sc ii} regions in Section 2. 
The oxygen abundances in the spiral galaxies NGC~3184, NGC~3351, NGC~5194, NGC~6744, 
NGC~2903, NGC~2997, NGC~5236 
and the Milky Way Galaxy are determined in Section 3. 
We discuss the reliability of the derived abundances 
and summarize our conclusions in Section 4.

For the line fluxes, we will be using the following notations throughout the paper: 
R$_2$ = $I_{[OII] \lambda 3727+ \lambda 3729} /I_{H\beta }$,
R$_3$ = $I_{[OIII] \lambda 4959+ \lambda 5007} /I_{H\beta }$, 
R = $I_{[OIII] \lambda 4363} /I_{H\beta }$, 
R$_{23}$ = R$_2$ + R$_3$. With these definitions,  
the excitation parameter P can then be expressed as: P = R$_3$/(R$_2$+R$_3$).

\section{Abundance derivation}

\subsection{Adopted equations for the T$_{\rm e}$ method}

A two-zone model for the temperature structure within the H\,{\sc ii} 
region was adopted. 
\citet{izotovetal05} have recently published a set 
of equations for the determination of the oxygen abundance 
in H\,{\sc ii} regions for a five-level atom. According to those authors,   
the electron temperature $t_3$ within the [O\,{\sc iii}] zone, in units of 10$^4$K, 
is given by the following equation 
\begin{eqnarray}
t_3 = \frac{1.432}{\log (R_{3}/R)  - \log C_{\rm T}}
\label{equation:t3}
\end{eqnarray}
where
\begin{equation}
C_{\rm T} = (8.44 - 1.09\,t_3 + 0.5\,t_3^2 - 0.08\,t_3^3) \, v 
\end{equation}
\begin{equation}
v = \frac{1 + 0.0004\,x_3}{1 + 0.044\,x_3}  
\end{equation}
and
\begin{equation}
x_3= 10^{-4} n_{\rm e} t_3^{-1/2}.
\end{equation}

As for the ionic oxygen abundances, they are derived from the following equations 
\begin{eqnarray}
12+ \log (O^{++}/H^+) = \log (I_{[OIII] \lambda 4959+ \lambda 5007}/I_{H_{\beta} })  +
  \nonumber  \\
6.200 + \frac{1.251}{t_3}  - 0.55 \log t_3 - 0.014\,t_3,
\label{equation:o3}
\end{eqnarray}
\begin{eqnarray}
12+ \log (O^{+}/H^+) = \log (I_{[OII] \lambda 3727+ \lambda 3729}/I_{ H_{\beta} })
 +  \nonumber  \\
5.961+ \frac{1.676}{t_2}  - 0.40 \log t_2  -0.034\,t_2 + \nonumber  \\
\log (1+1.35x_2)  .
\label{equation:o2}
\end{eqnarray}

\begin{equation}
x_2= 10^{-4} n_{\rm e} t_2^{-1/2}.
\end{equation}
Here $n_e$ is the electron density in cm$^{-3}$. 

The total oxygen abundances are then derived from the following equation 
\begin{equation}
\frac{O}{H} = \frac{O^+}{H^+} + \frac{O^{++}}{H^+}                .
\label{equation:otot}
\end{equation}

The electron temperature $t_2$ of the [O\,{\sc ii}] zone is 
usually determined from an equation which relates $t_2$ to $t_3$, 
derived by fitting H\,{\sc ii} region models. 
Several versions of this $t_2$ -- $t_3$ relation have been proposed. 
A widely used relation has been suggested by \citet{campbelletal86} 
(see also \citet{garnett92}) based on the H\,{\sc ii} region models of 
\citet{stasinska82}. \citet{campbelletal86} has found that the $t_2$ -- $t_3$  
relationship can be parameterized as
\begin{equation}
t_2 = 0.7 \, t_3 + 0.3.
\label{equation:ttgar}
\end{equation}
It will be used here.

\subsection{The ff -- relation}

The fluxes R in the auroral lines are necessary to derive the oxygen abundances 
in H\,{\sc ii} regions using the T$_{\rm e}$ method. 
But they are faint and often undetectable  
in spectra of high-metallicity H\,{\sc ii} regions. 
It was shown \citep{ff} that the flux R in the auroral line is related to the 
total flux R$_{23}$ in the strong nebular lines through a relation of the 
type
\begin{equation}
\log R = a + b \times \log R_{23}.
\label{equation:ffe}
\end{equation}
Eq.(\ref{equation:ffe}) will be hereinafter referred to as the 
flux -- flux or ff -- relation. It was 
found that this relationship is metallicity-dependent at low metallicities, 
but becomes independent of metallicity (within the uncertainties of 
the available data) at metallicities higher than 12+logO/H $\sim$ 8.25, 
i.e. there is one-to-one correspondence 
between the auroral and nebular oxygen line fluxes  
in spectra of high-metallicity H\,{\sc ii} regions. 
\citet{ff} derived
\begin{equation}
\log R = -4.264 + 3.087 \, \log R_{23}.
\label{equation:ff}
\end{equation}
We can now make use of the ff -- relation to define a ``discrepancy index'',
equal to the  difference between the logarithm of the 
observed flux R$^{{\rm obs}}$ in the [O\,{\sc iii}]$\lambda$4363 line 
and that of the flux R$^{\rm cal}$ of that line derived from the strong  
[O\,{\sc ii}]$\lambda$3727, [O\,{\sc iii}]$\lambda$$\lambda$4959,5007 
lines using the ff - relation: 
\begin{equation}
D_{{\rm ff}} =  \log R^{{\rm obs}} - \log R^{\rm cal}.
\label{equation:d}
\end{equation}

Since R$_{23}$ = R$_3$/P, the 
ff -- relation can be also expressed in the form R=f(R$_3$,P). 
We will consider an expression of the type 
\begin{equation}
\log R = a_1 + a_2 \, \log P + a_3 \, \log R_{3} + a_4 \, (\log P)^2 .
\label{equation:ffpe}
\end{equation}

\begin{figure}
\resizebox{1.00\hsize}{!}{\includegraphics[angle=000]{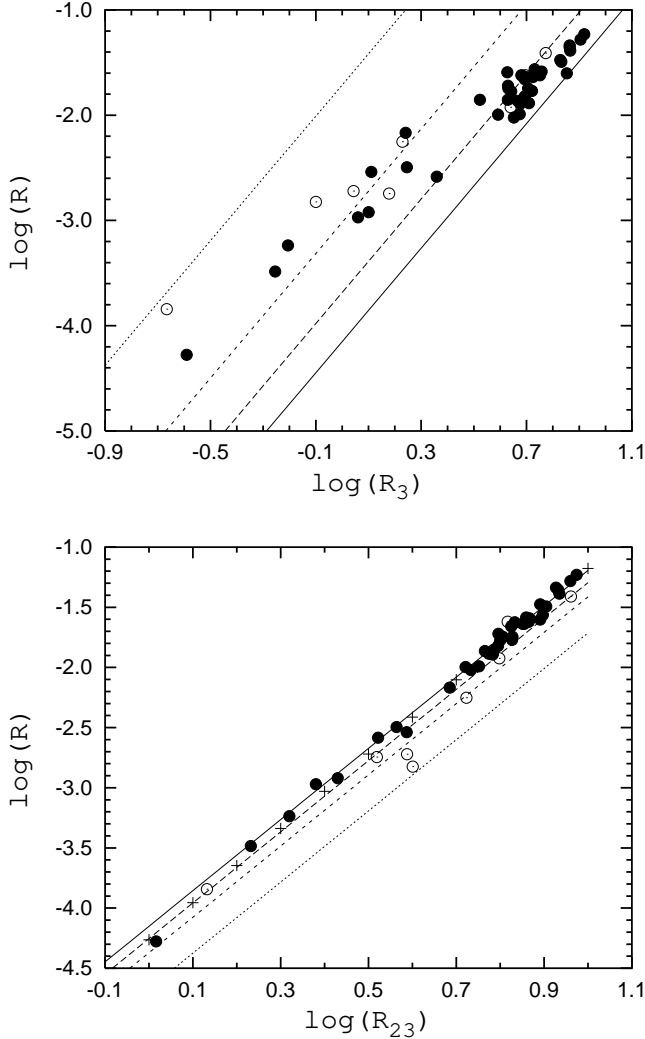}}
\caption{
The ff -- relations for H\,{\sc ii} regions. 
{\it Top panel.} Flux R in the oxygen auroral line as a function of the 
flux R$_{3}$ in the strong nebular [O\,{\sc iii}]$ \lambda 4959+ \lambda 5007$ 
lines for H\,{\sc ii} regions. 
The circles (open and filled) show H\,{\sc ii} regions with 
12+logO/H $>$ 8.25. 
The relations corresponding to Eq.(\ref{equation:ffp}) for different values 
of the excitation parameter are shown by the solid (P = 1.0),  
long-dashed  (P = 0.7), short-dashed  (P = 0.3), 
and dotted (P = 0.1) lines. 
Only filled circles are used in deriving Eq.(\ref{equation:ffp}). 
{\it Bottom panel.} 
Flux R in the oxygen auroral line as a function of the total flux R$_{23}$ 
in the strong 
nebular [O\,{\sc ii}]$ \lambda 3727+ \lambda 3729$ and 
[O\,{\sc iii}]$ \lambda 4959+ \lambda 5007$ lines. 
The open and filled circles have the same meaning as in the top panel. 
The relations corresponding to Eq.(\ref{equation:ffp}) for different values 
of the excitation parameter are shown by the solid  (P = 1.0), 
long-dashed  (P = 0.3), short-dashed  (P = 0.2), 
and dotted  (P = 0.1) lines. 
The relation corresponding to Eq.(\ref{equation:ff}) is shown by plus signs. 
}
\label{figure:f-f}
\end{figure}

\begin{figure}
\resizebox{1.00\hsize}{!}{\includegraphics[angle=000]{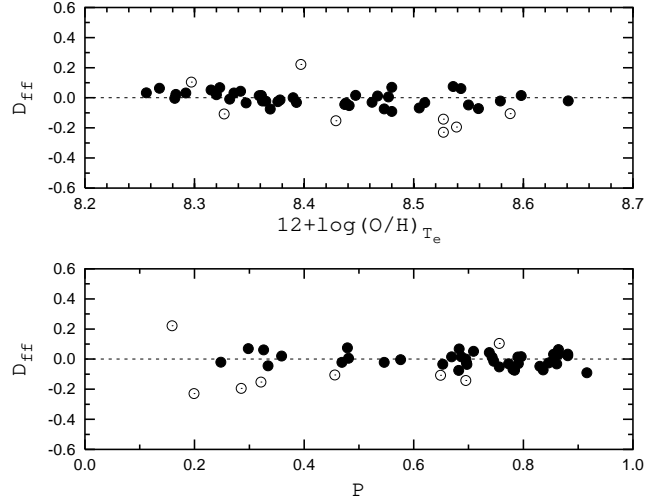}}
\caption{Deviations from the ff -- relation as parameterized by 
the discrepancy index 
D$_{{\rm ff}}$ as a function of 
oxygen abundance ({\it top panel}) and of excitation parameter P 
({\it bottom panel}). 
The filled circles are H\,{\sc ii} regions used in deriving the ff -- relation. 
The open circles are other H\,{\sc ii} regions. 
}
\label{figure:f-df}
\end{figure}

Using a sample of H\,{\sc ii} regions with recent high-precision measurements 
of oxygen lines fluxes, the values of the coefficients in 
Eq.(\ref{equation:ffpe}) can be derived. 
Such a sample has been compiled by Pilyugin 
(2005). 
\citet{bresetal04} have recently detected the 
auroral lines [S\,{\sc iii}]$\lambda$6312 and/or [N\,{\sc ii}]$\lambda$5755 
and determined T$_e$-based abundances in 10 faint H\,{\sc ii} regions 
in the spiral galaxy M~51. Unfortunately they were unable to detect 
the oxygen auroral line. 
Using the electron temperature t$_3$ recommended by \citet{bresetal04}
and the equations given above for the T$_e$-method, we have estimated the value of R which 
corresponds to the t$_3$ for every H\,{\sc ii} region. 
It has been found \citep{ff} that the oxygen line fluxes in six 
faint H\,{\sc ii} regions in the spiral galaxy M~51 (CCM~54, CCM~55, CCM~57, CCM~57A, CCM~71A, and 
CCM~84A) satisfy the ff -- relation. 
These data are included in the present sample to enlarge the ranges in P and 
R$_3$. Thus, our sample consists of a total of 48 data points.

The values of the coefficients in Eq.(\ref{equation:ffpe}) 
are derived by using an iteration procedure. In the first step, the relation is 
determined from all data using the least-square method. Then, the point with  
the largest deviation is rejected, and a new relation is derived. The iteration 
procedure is pursued until two successive relations have  
all their coefficients differing by less 0.001 and the absolute value of the largest 
deviation is less than 0.1 dex. 
The following ff -- relation was obtained
\begin{eqnarray}
\log R & = & - 4.151- 3.118\,\log P + 2.958 \, \log R_{3}
\nonumber  \\
       & - & 0.680 \, (\log P)^2 .
\label{equation:ffp}
\end{eqnarray}

\subsection{Characteristics of the ff -- relation}

The top panel in Fig.~\ref{figure:f-f} shows the flux R in the oxygen auroral 
line as a function of the flux R$_{3}$ in strong nebular 
[O\,{\sc iii}]$ \lambda 4959+ \lambda 5007$ lines. 
The open and filled circles show individual H\,{\sc ii} regions with 
12+logO/H $>$ 8.25. 
The relations corresponding to Eq.(\ref{equation:ffp}) for different values 
of the excitation parameter are shown by the solid (P = 1.0),
long-dashed  (P = 0.7), short-dashed (P = 0.3), 
and dotted (P = 0.1) lines. 
The filled circles show the data used in deriving Eq.(\ref{equation:ffp})
(40 out of 48 original data points).

The deviations from the ff -- relation given by Eq.(\ref{equation:ffp}) 
as parameterized by the discrepancy index D$_{{\rm ff}}$ are shown in 
Fig.~\ref{figure:f-df} as a function of oxygen abundance (top panel) 
and of excitation parameter P (bottom panel). 
The filled circles are H\,{\sc ii} regions used 
in deriving Eq.(\ref{equation:ffp}). 
The open circles are the other H\,{\sc ii} regions. 
Fig.~\ref{figure:f-df} shows that the deviations  
do not show a correlation either with 
metallicity or with the excitation parameter. 

Let us compare the two ff - relations given by 
Eq.(\ref{equation:ff}) and Eq.(\ref{equation:ffp}). 
The bottom panel in Fig.~\ref{figure:f-f} shows the   
flux R in the oxygen auroral line as a function of the total flux R$_{23}$ in strong 
nebular [O\,{\sc ii}]$ \lambda 3727+ \lambda 3729$ and 
[O\,{\sc iii}]$ \lambda 4959+ \lambda 5007$ lines. 
The filled circles are the H\,{\sc ii} regions used 
in deriving Eq.(\ref{equation:ffp}). 
The open circles are the other H\,{\sc ii} regions. 
The relations corresponding to Eq.(\ref{equation:ffp}) for different values 
of the excitation parameter are shown by the solid  (P = 1.0), 
long-dashed  (P = 0.3), short-dashed  (P = 0.2), 
and dotted (P = 0.1) lines. 
The relation corresponding to Eq.(\ref{equation:ff}) is shown by the plus signs. 
Inspection of the bottom panel of Fig.~\ref{figure:f-f} shows that  
the relations given by Eq.(\ref{equation:ffp}) for values 
of the excitation parameter P ranging from 1 to $\sim$ 0.3 are close 
each to other. That led \citet{ff} to conclude that 
there appears to be no third parameter in the ff -- relation. 
Indeed, examination of the bottom panel in Fig.~\ref{figure:f-f} shows  
that  the ff -- relations given by 
Eq.(\ref{equation:ff}) and Eq.(\ref{equation:ffp}) are very similar to each  
other and reproduce the observational data well at high values of the 
excitation parameter, P $\ga $ 0.3. At the same time there is 
an appreciable divergence between those two relations  
at low values 
of the excitation parameter,  P $\la $ 0.3. 
For definiteness's sake, we will use the ff -- relation given by 
Eq.(\ref{equation:ffp}) for all values of P. 

It should be noted however that the particular form of the analytical 
expression adopted for the ff -- relation may be questioned. We have 
chosen a simple form, Eq.(\ref{equation:ffpe}),  
but perhaps a more complex expression may give a better fit to   
the auroral -- nebular oxygen line fluxes relationship. Furthermore, the 
coefficients in the adopted expression are derived using calibrating   
H\,{\sc ii} regions which have a discrepancy index D$_{{\rm ff}}$  
as large as 0.1 dex in absolute value. Perhaps smaller absolute values of 
D$_{{\rm ff}}$  may result in a more precise relation.
It can be seen that the calibration curves give a satisfactory fit to 
the  observational data. At the same time, Fig.~\ref{figure:f-f} shows that 
the available  measurements for calibrating faint H\,{\sc ii} regions 
(those with  low R$_3$) are very few in number. 
Furthermore, the R values for the faint H\,{\sc ii} regions from \citet{bresetal04} 
are not measured but estimated from the electron 
temperatures t$_3$ and, consequently, they are not beyond question. Clearly, 
high-precision measurements of oxygen lines fluxes  
in faint H\,{\sc ii} regions are needed to check  
the derived ff -- relation.

\subsection{The (O/H)$_{\rm ff}$ abundances}

\begin{figure}
\resizebox{1.00\hsize}{!}{\includegraphics[angle=000]{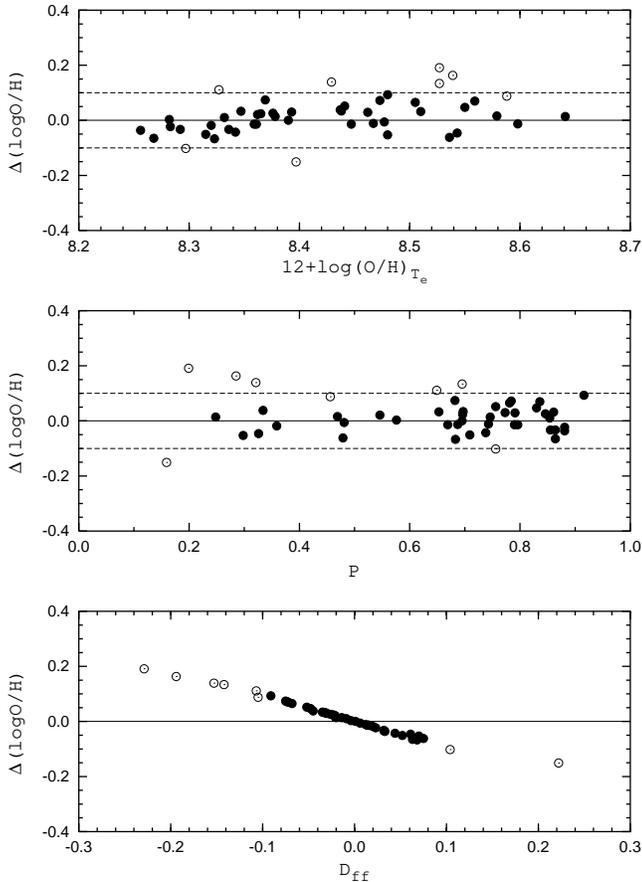}}
\caption{The difference $\Delta$log(O/H) = log(O/H)$_{{\rm T_e}}$ --  
log(O/H)$_{{\rm ff}}$ as a function of log(O/H)$_{{\rm T_e}}$ 
({\it top panel}), of excitation parameter P ({\it middle panel}), 
and of discrepancy index D$_{{\rm ff}}$ ({\it bottom panel}). 
}
\label{figure:z-dz}
\end{figure}

Using the T$_{\rm e}$ method, both (O/H)$_{\rm T_e}$ abundances 
based on the measured line fluxes R$^{\rm obs}$, and (O/H)$_{\rm ff}$ abundances 
based on the line fluxes R$^{\rm cal}$ determined from the ff -- relation, can be 
derived for our sample of H\,{\sc ii} regions. 
Comparison of their values gives us a check on  
 the (O/H)$_{\rm ff}$ abundances. 
Fig.~\ref{figure:z-dz} shows the 
difference $\Delta$log(O/H) = log(O/H)$_{{\rm T_e}}$ --  
log(O/H)$_{{\rm ff}}$ as a function of log(O/H)$_{{\rm T_e}}$ 
(top panel), of excitation parameter P (middle panel), 
and of discrepancy index D$_{{\rm ff}}$ (bottom panel). 
Inspection of
Fig.~\ref{figure:z-dz} shows that the differences $\Delta$log(O/H) 
do not show an appreciable correlation with 
metallicity or with excitation parameter. There is however an anticorrelation  
of $\Delta$log(O/H) with D$_{{\rm ff}}$. 
This can be easily understood. 
The discrepancy index appears to be an indicator of the error 
in the auroral line R measurements. In that case, positive values of 
D$_{{\rm ff}}$ imply overestimated R fluxes, which result in turn in overestimated  
electron temperatures and underestimated oxygen abundances, i.e. negative $\Delta$log(O/H),
just the observed trend.
In general, the (O/H)$_{{\rm ff}}$ abundances agree satisfactorily with 
the (O/H)$_{{\rm T_e}}$ abundances and 
do not show any systematic trend.

In summary, the flux in the faint auroral line [O\,{\sc iii}]$\lambda 4363$ 
can be estimated reasonably well from the measured fluxes in the strong nebular lines 
[O\,{\sc ii}]$ \lambda 3727+ \lambda 3729$ and 
[O\,{\sc iii}]$ \lambda 4959+ \lambda 5007$ via the ff -- relation. 
Then, using the derived flux in the auroral line, the oxygen abundance 
can be found through the classic T$_{\rm e}$ -- method.

\section{Oxygen abundances in spiral galaxies}

Here the derived ff -- relation will be applied to derive the oxygen abundances  
in a number of spiral galaxies. 
First, M~51 (NGC~5194), 
the oxygen-richest spiral galaxy ((O/H)$_{\rm C}$ = 9.54) in the 
sample of \citet{vila92} will be considered. The value of (O/H)$_{\rm C}$ in 
M~51   was derived recently from 
determinations of (O/H)$_{\rm T_e}$ abundances in a number of H\,{\sc ii} 
regions by  \citet{bresetal04}. 
Comparison between the radial distributions of (O/H)$_{\rm T_e}$ and 
(O/H)$_{\rm ff}$ abundances in the disk of M~51 provides another 
possibility to test the reliability of the latter.    
Second, (O/H)$_{\rm ff}$ abundances  in  
NGC~3184, NGC~3351, and  NGC~6744 which have been reported to be the most oxygen-rich 
spirals, will be determined.  
Third, we will consider three well-observed spiral galaxies 
NGC~2903, NGC~2997, and  NGC~5236 for comparison.     
Fourth, the ff -- relation will be applied to Galactic H\,{\sc ii} regions 
to select high-precision measurements.

\subsection{Oxygen abundances in  M~51}

\begin{figure}
\resizebox{1.00\hsize}{!}{\includegraphics[angle=000]{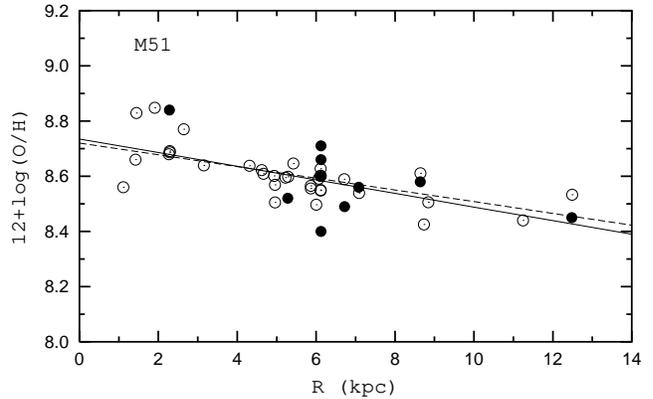}}
\caption{Radial distribution of oxygen abundances 
in the disk of the spiral galaxy  M~51 (NGC~5194). 
The open circles denote (O/H)$_{\rm ff}$ abundances. 
The solid line is the linear least-square 
best fit to those data.
The filled circles show (O/H)$_{\rm T_e}$ abundances from \citet{bresetal04}. 
The dashed line is their linear regression fit. 
}
\label{figure:m51}
\end{figure}

The compilation of published spectra of H\,{\sc ii} regions in 
M~51 is taken from \citet{pilvilcon04}. 
The H\,{\sc ii} regions with measured R$_3$ and R$_2$ fluxes from 
\citet{bresetal04} (their Tables 1 and 6) have been added. 
We determine (O/H)$_{\rm ff}$ abundances for 
every H\,{\sc ii} region using the derived ff -- relation. 
The (O/H)$_{{\rm ff}}$ abundances in the H\,{\sc ii} regions of  
M~51 are shown as a function of galactocentric distance in 
Fig.~\ref{figure:m51} by open circles. 
The solid line is the linear least-square best fit to those data:  
\begin{equation}
12+\log(O/H) = 8.74\,(\pm 0.03) - 0.025\, (\pm 0.004) \times R .
\label{equation:m51f}
\end{equation}
The distance of M~51 (d = 7.64 Mpc) and the isophotal radius 
(R$_{25}$ = 5.61 arcmin) were taken from \citet{pilvilcon04}. 
The filled circles are 
(O/H)$_{\rm T_e}$ abundances determined by \citet{bresetal04}. 
The dashed line is their linear regression: 
\begin{equation}
12+\log(O/H) = 8.72\,(\pm 0.09) - 0.02 \, (\pm 0.01) \times R .
\label{equation:m51gar}
\end{equation}

Fig.~\ref{figure:m51} shows that the radial distribution of (O/H)$_{{\rm ff}}$ 
abundances agrees well with that of the (O/H)$_{\rm T_e}$ abundances 
(compare also Eq.(\ref{equation:m51f}) and Eq.(\ref{equation:m51gar})).  
This is a strong argument in favor of the reliability of the (O/H)$_{{\rm ff}}$ 
abundances.

Some H\,{\sc ii} regions show large deviations from the 
general radial trend in oxygen abundances. Some of those deviations 
can be real, but some are probably caused by uncertainties 
in the oxygen abundance determinations. It should be emphasized  
that the uncertainties in the (O/H)$_{{\rm ff}}$ abundances are not necessarily  
caused by uncertainties in the line flux measurements.
It has been noted \citep{ff} that the ff -- relation gives reliable results only if two 
conditions are satisfied; {\it i}) the measured fluxes reflect their 
relative contributions to the radiation of the whole nebula, and  
{\it ii}) the H\,{\sc ii} region is ionization-bounded. 
If these two conditions are not met, then the derived (O/H)$_{{\rm ff}}$  
abundances may be significantly in error.

\begin{figure}
\resizebox{1.00\hsize}{!}{\includegraphics[angle=000]{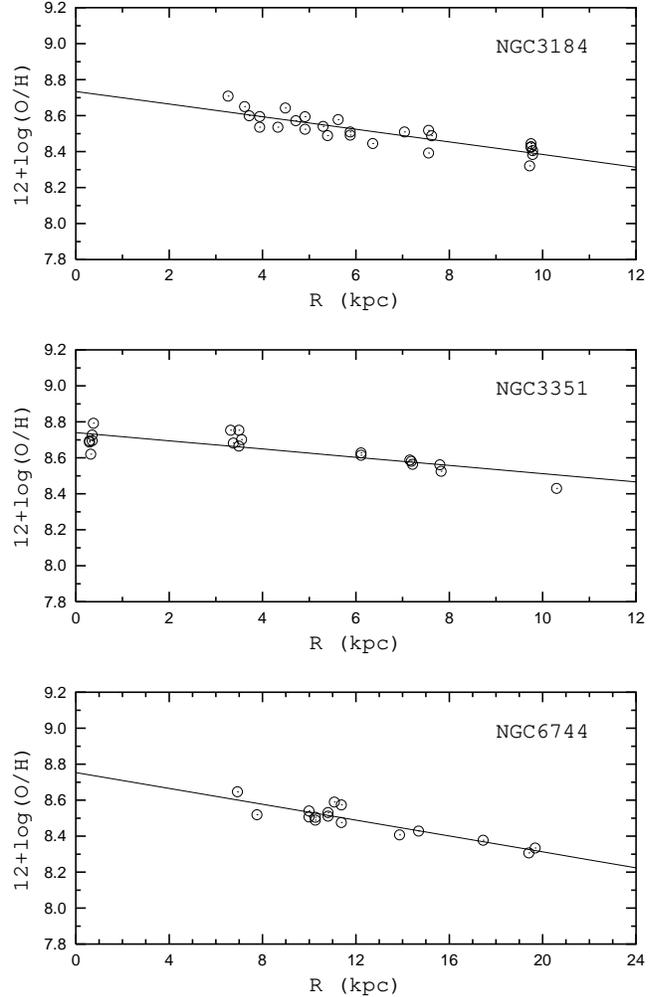}}
\caption{Radial distributions of (O/H)$_{\rm ff}$ abundances 
in the spiral galaxies NGC~3184, NGC~3351, and NGC~6744. 
The open circles show individual H\,{\sc ii} regions. 
The lines are the least-square fits to those data. 
}
\label{figure:spmax}
\end{figure}

\begin{figure}
\resizebox{1.00\hsize}{!}{\includegraphics[angle=000]{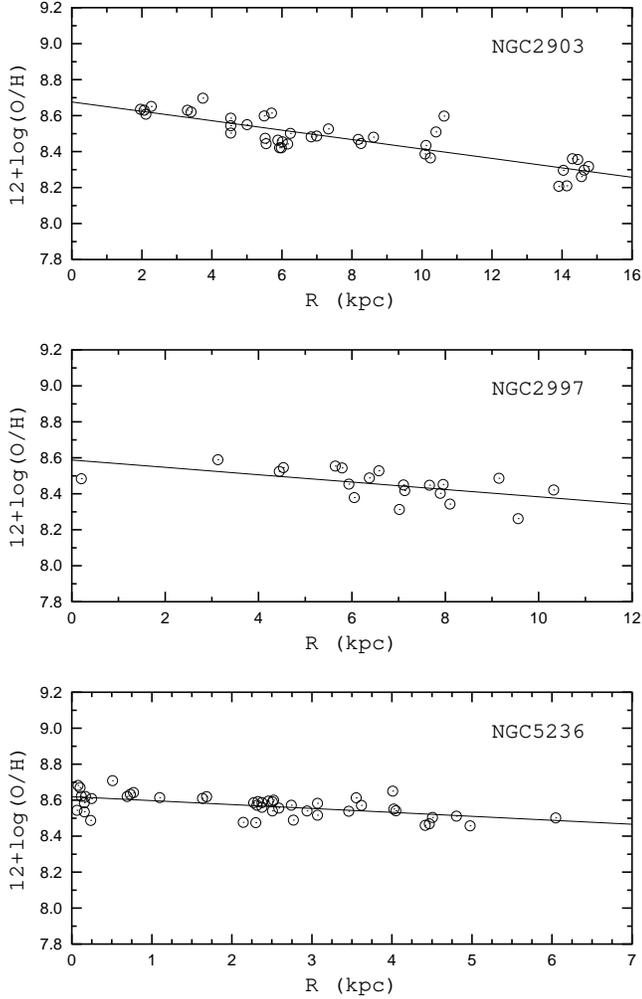}}
\caption{Radial distributions of (O/H)$_{\rm ff}$ abundances 
in the spiral galaxies NGC~2903, NGC~2997, and NGC~5236. 
The open circles show individual H\,{\sc ii} regions. 
The lines are the least-square fits to those data.
}
\label{figure:splow}
\end{figure}

\subsection{Oxygen abundances in the most oxygen-rich spiral galaxies}

The compilation of published spectra of H\,{\sc ii} regions in the spiral galaxies 
NGC~3184, NGC~3351, and  NGC~6744 was taken from \citet{pilvilcon04}. 
The (O/H)$_{{\rm ff}}$ abundances in the H\,{\sc ii} regions of 
NGC~3184 is shown as a function of galactocentric distance in 
the top panel of Fig.~\ref{figure:spmax}. The solid line is the linear least-square best fit 
to those data: 
\begin{equation}
12+\log(O/H) = 8.73\,(\pm 0.03) - 0.035 \, (\pm 0.004) \times R .
\label{equation:n3184}
\end{equation}
The same data are shown for the H\,{\sc ii} regions of NGC~3351 in 
the middle panel of Fig.~\ref{figure:spmax}. The solid line is the linear least-square best fit 
to those data:   
\begin{equation}
12+\log(O/H) = 8.74\,(\pm 0.02) - 0.023 \, (\pm 0.004) \times R .
\label{equation:n3351}
\end{equation}
Similarly, the data for  the H\,{\sc ii} regions of 
NGC~6744 are shown in 
the bottom panel of Fig.~\ref{figure:spmax}. The linear least-square best fit 
to those data is given by:  
\begin{equation}
12+\log(O/H) = 8.75\,(\pm 0.04) - 0.022 \, (\pm 0.003) \times R .
\label{equation:n6744}
\end{equation}
The distances of the galaxies and the 
galactocentric distances of the H\,{\sc ii} regions were taken 
from \citet{pilvilcon04}. 

Figs.~\ref{figure:m51} and ~\ref{figure:spmax} (see also 
Eqs.(\ref{equation:m51f},\ref{equation:n3184},
\ref{equation:n3351},\ref{equation:n6744}))
show that the values of (O/H)$_{\rm C}$ in these spiral galaxies 
 are $\sim$ 8.75.

\subsection{Oxygen abundances in other spiral galaxies}

We will consider here three well-observed spiral galaxies for comparison.
The compilation of the published spectra of H\,{\sc ii} regions in the spiral galaxies 
NGC~2903, NGC~2997, and  NGC~5236 was taken from \citet{pilvilcon04}. 
Recent measurements 
from \citet{bresstas05} have been added. 
Oxygen abundances were calculated for 
every H\,{\sc ii} region  using the derived ff -- relation.

The (O/H)$_{{\rm ff}}$ abundances in the H\,{\sc ii} regions of  
NGC~2903 are shown as a function of galactocentric distance in 
the top panel of Fig.~\ref{figure:splow}. The solid line is the linear least-square 
best fit to those data:  
\begin{equation}
12+\log(O/H) = 8.68\,(\pm 0.02) - 0.026 \, (\pm 0.003) \times R .
\label{equation:n2903}
\end{equation}
The middle panel of Fig.~\ref{figure:splow} shows the same for  
NGC~2997. 
The linear least-square fit to those data is:    
\begin{equation}
12+\log(O/H) = 8.59\,(\pm 0.05) - 0.021 \, (\pm 0.007) \times R .
\label{equation:n2997}
\end{equation}
The data for NGC~5236 are shown  in 
the bottom panel of Fig.~\ref{figure:splow}. The linear 
least-square fit to those data is:  
\begin{equation}
12+\log(O/H) = 8.62\,(\pm 0.01) - 0.022 \, (\pm 0.005) \times R .
\label{equation:n5236}
\end{equation}
The distances of the galaxies 
 were taken from \citet{pilvilcon04}. 

Fig.~\ref{figure:splow} (see also 
Eqs.(\ref{equation:n2903},\ref{equation:n2997},\ref{equation:n5236}))
shows that the (O/H)$_{\rm C}$ abundances in  
NGC~2903, NGC~2997, and NGC~5236 are smaller than 8.70, i.e. lower than in
NGC~3184, NGC~3351, NGC~5194, and NGC~6744. 

Thus, the maximum gas-phase oxygen abundance in the oxygen-richest spiral 
galaxies is 12+log(O/H) $\sim$ 8.75. 
This value is significantly lower (by a factor of 5 or more) than the previous
value based on the one-dimensional empirical 
calibrations \citep[e.g.][]{vila92,zkh,vanzeeetal98,garnett02}.

\subsection{Oxygen abundances in the Milky Way Galaxy}

\begin{figure}
\resizebox{1.00\hsize}{!}{\includegraphics[angle=000]{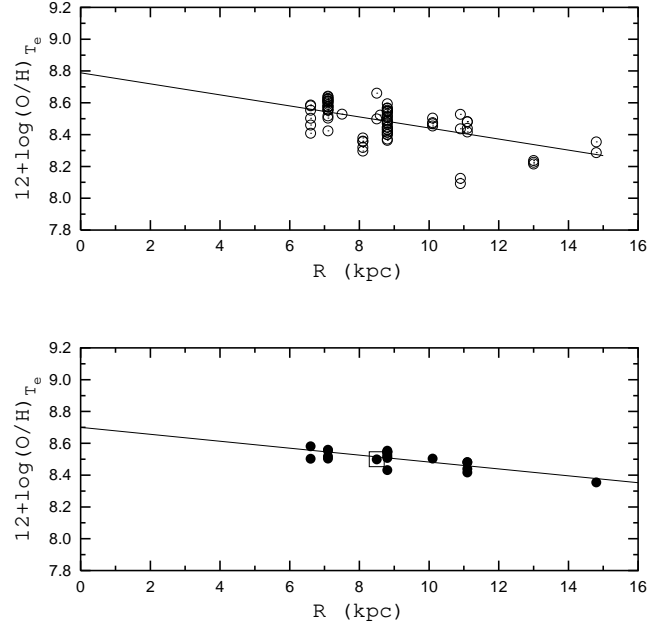}}
\caption{
Radial distribution of (O/H)$_{{\rm T_e}}$ abundances in 
the disk of the Milky Way Galaxy. {\it Top panel.} 
The open circles show (O/H)$_{\rm T_e}$ for all H\,{\sc ii} regions. 
The solid line is the linear least-square fit to those data. 
{\it Bottom panel.}  The filled circles show (O/H)$_{\rm T_e}$ 
abundances for H\,{\sc ii} regions with an absolute value of 
the discrepancy index D$_{\rm ff}$ less than 0.05 dex. 
The solid line is the linear least-square fit to those data. 
The open square shows the 
oxygen abundance in the interstellar gas at the solar galactocentric distance.
}
\label{figure:mwg}
\end{figure}

We now consider the Milky Way Galaxy. 
A compilation of published spectra of Galactic H\,{\sc ii} regions with 
an auroral [O\,{\sc iii}]$\lambda$4363 line has been carried 
out by Pilyugin et al. (2003). That list contains 69 individual measurements of 
13 H\,{\sc ii} regions in the range of galactocentric distances from 6.6 to 
14.8 kpc. Recent measurements of six Galactic H\,{\sc ii} regions by 
\citet{estebanetal05} were added to that list, resulting in a total of 75  
measurements. The T$_{\rm e}$-based oxygen abundances were estimated using the  
equations above. The derived (O/H)$_{{\rm T_e}}$ abundances are shown as a 
function of 
galactocentric distance by the open circles in the top panel of Fig.~\ref{figure:mwg}. 
The linear least-square best fit to those data (with two points with a 
deviation more than 0.2 dex rejected)
\begin{equation}
12+\log(O/H) = 8.79\,(\pm 0.05) - 0.034 \, (\pm 0.005) \times R 
\label{equation:mwgte0}
\end{equation}
is shown by the solid line in the top 
panel of Fig.~\ref{figure:mwg}.

There is a relative large number of measurements of abundances
in the Milky Way Galaxy. They show a large scatter since 
the quality of the data obtained over a period of more than thirty years 
is necessarily heterogeneous. Some measurements   
have a low accuracy. 
The ff -- relation provides a way to select out the H\,{\sc ii} regions 
with high quality measurements \citep{pilyuginthuan05}. 
The bottom panel of Fig.~\ref{figure:mwg} shows the (O/H)$_{\rm T_e}$ abundances 
as a function of galactocentric distance for objects with an absolute 
value of the discrepancy index D$_{\rm ff}$ less than 0.05 dex. 
The linear least-square best fit to those data
\begin{equation}
12+\log(O/H) = 8.70\,(\pm 0.04) - 0.022 \, (\pm 0.004) \times R 
\label{equation:mwgte}
\end{equation}
is shown by the solid line in the bottom panel of Fig.~\ref{figure:mwg}. 
The open square is the 
oxygen abundance of the interstellar gas at the solar galactocentric distance
(see below).

The central oxygen abundance in the Milky Way obtained here, 
(O/H)$_{\rm C}$ $\sim$ 8.70, is significanly lower than the widely used  
(O/H)$_{\rm C}$ = 9.38 of 
\citet{shaveretal83}.
Our oxygen abundance radial gradient is in agreement with that of \citet{dafloncunha04}.
They derived the relation
\begin{equation}
12+\log(O/H) = 8.762\,(\pm 0.105) - 0.031 \, (\pm 0.012) \times R 
\label{equation:star}
\end{equation}
from abundance determinations in young OB stars.

We note that the  central oxygen abundance obtained here for  
the Milky Way is close to that in other large spirals.

\section{Discussion and conclusions} 

The (O/H)$_{\rm ff}$ abundances determined here rely on the validity of the 
classic T$_e$ -- method. This has been questioned for the high-metallicity 
regime in a number of investigations by 
comparison with H\,{\sc ii} region photoionization models. One 
should keep in mind however that the existing numerical models of H\,{\sc ii} 
regions are far from being perfect \citep{stasinska04}, and, hence, the 
statement that H\,{\sc ii} region models provide more realistic abundances as 
compared to the T$_e$ -- method should not be taken for granted \citep{vybor}. 
Recently \citet{stasinska05} has examined the biases involved in 
T$_{\rm e}$-based 
abundance determinations at high metallicities. 
She has found that, as long as the metallicity is low, the derived 
(O/H)$_{\rm T_e}$ value is very close to the real one. Discrepancies
appear around 12+log(O/H) = 8.6 - 8.7, and may become very large 
as the metallicity 
increases \citep[Fig. 1a in][]{stasinska05}. The derived (O/H)$_{\rm T_e}$ 
values are smaller than the real ones, sometimes by enormous factors. This 
means that, if such metal-rich H\,{\sc ii} regions exist, the classic 
T$_e$ -- method will always lead to systematically lower 
derived oxygen abundances.

Can the effect discussed by Stasi\'{n}ska be responsible for the low 
(O/H)$_{\rm T_e}$  abundances obtained here? 
Let us assume that it is the case, 
and that 
the true central oxygen abundances in the spiral galaxies are higher than
12 + log(O/H) = 9.0. What radial distributions of (O/H)$_{\rm T_e}$  
abundances 
in the disk of spiral galaxies would one expect in this case? 
Starting from the periphery of a galaxy, the (O/H)$_{\rm T_e}$  abundances 
would increase with 
decreasing galactocentric distance (following the true abundances) until 
the (O/H)$_{\rm T_e}$  abundance reaches 
the value of 12+log(O/H) $\sim$ 8.7. 
After that the (O/H)$_{\rm T_e}$ abundances 
would decrease with decreasing galactocentric distance because of the 
Stasi\'{n}ska \citep[Fig. 1a in][]{stasinska05} effect, 
although the true abundances continue to 
increase with decreasing galactocentric distance. 
Thus, the radial distribution of (O/H)$_{\rm T_e}$  abundances should show a
bow-shaped curve with a maximum value of 12+log(O/H) $\sim$ 8.7 
at some galactocentric distance. 

Do the actual data (Fig. 5--7) show such a behavior? 
The derived radial distributions of 
(O/H)$_{\rm T_e}$  abundances in the disks of spiral galaxies do not show an 
appreciable curvature, and the (O/H)$_{\rm T_e}$  abundances 
increase more or less monotonically with decreasing galactocentric distance. 
Only the H\,{\sc ii} regions in the very central parts of the spiral galaxies 
NGC~3351 (the middle panel of Fig.~\ref{figure:spmax}) and NGC~2997 
(the middle panel of Fig.~\ref{figure:splow}) may show (O/H)$_{\rm T_e}$  
abundances which are slightly lower than expected from  
the general radial abundance trends, but the effect is very slight and may not 
be real if errors of 0.1 dex (Fig. 3) in the calibration 
and in the line intensity measurements are taken into account. 
To settle the question of whether the abundances in the metal-richest 
H\,{\sc ii} regions 
are affected by the Stasi\'{n}ska's effect, many 
more accurate measurements of H\,{\sc ii} regions (including 
those in the very central parts) in the most 
oxygen-rich spiral galaxies are necessary. 

Thus, while our results do not rule out 
the possible existence of the Stasi\'{n}ska's 
effect, they suggest that the great majority of 
H\,{\sc ii} regions in galaxies are in a  
metallicity range where this effect is not important. Indeed, the 
Stasi\'{n}ska's effect becomes important only 
at oxygen abundances higher than 
12 + log(O/H) $\sim$ 8.7. We have found that this "critical" value 
is reached only in the central part of the most oxygen-rich galaxies. 
We conclude that, in the light of our present results,
H\,{\sc ii} regions with 
12+log(O/H) $>$ 8.7 are rare, if they exist at all. It should  
be noted that 
Stasi\'{n}ska also questioned the existence of such metal-rich 
H\,{\sc ii} regions \citep[Section 2.2 in][]{stasinska05}. 

Another factor that can affect T$_{\rm e}$-based abundance determinations is 
the temperature fluctuations inside H\,{\sc ii} regions \citep{peimbert67}. 
If they are important, the (O/H)$_{\rm T_e}$ abundance would be a lower limit.
 In most cases the 
effect appears to be small, probably under 0.1 dex \citep{pagel03}.

Fortunately, there is a way to verify the validity of the T$_{\rm e}$ -- method at 
high metallicities  \citep{vybor}. 
High-resolution observations of the weak interstellar O{\sc i}$\lambda$1356 
absorption lines towards stars allow one to determine the interstellar oxygen 
abundance in the solar vicinity with a very high precision. It should be noted 
that this method is model-independent.  These observations 
yield a mean interstellar oxygen abundance of 284 -- 390 O atoms per 
10$^6$ H atoms (or 12+log(O/H) = 8.45 -- 8.59) 
\citep{meyeretal98,sofiameyer01,cartledgeetal04}.
  \citet{oliveiraetal05} have determined a mean O/H ratio 
of 345 $\pm$19 O atoms per 10$^6$ H atoms (or 12+log(O/H) = 8.54) 
for the Local Bubble. Thus, an oxygen abundance 12+log(O/H) = 8.50  
in the interstellar gas at the solar galactocentric distance seems 
to be a reasonable value.
The value of the oxygen abundance at the solar galactocentric distance 
traced by (O/H)$_{\rm T_e}$ abundances in H\,{\sc ii} regions is then in 
good agreement with the oxygen abundance derived with high precision from the 
interstellar absorption lines towards stars 
(open square in the bottom panel of Fig.~\ref{figure:mwg}). 
 This is strong evidence that 
the classic T$_{\rm e}$ -- method provides accurate oxygen abundances in 
high-metallicity H\,{\sc ii} regions. 

The measured oxygen abundance in the solar vicinity provides an indirect 
way to check our derived central oxygen abundances. 
The oxygen abundance in the solar vicinity is 12+log(O/H)=8.50, and a  
gas mass fraction $\mu$ $\sim$ 0.20 seems appropriate.
The simple model of chemical evolution of galaxies predicts that a 
decrease of $\mu$ by 0.1 results in a increase of the oxygen
abundance by $\sim$ 0.12 dex in the range of $\mu$ from $\sim$ 0.75 to 
$\sim$ 0.05. 
When all the gas in the solar vicinity will be converted 
into stars, the oxygen abundance will increase by 0.2 -- 0.25 dex and 
will reach a value around 12+log(O/H) = 8.70 -- 8.75. This value is
in excellent agreement with the central oxygen abundances obtained above 
for the Milky Way and other galaxies. 
There is no room for central oxygen 
abundances as large as 12+logO/H = 9.00 or greater. 
 
The present study suggests that there is an upper limit 
to the oxygen abundances in spiral galaxies. If true, the existing luminosity 
-- metallicity relation should be revisited. One can expect 
the metallicity to increase with luminosity up to 12+(O/H) $\sim$ 8.75, 
but to remain approximately constant for higher metallicities, 
resulting in a flattening of 
the luminosity -- metallicity relation. This suggestion will be 
investigated in a future paper.  

The maximum gas-phase oxygen abundance in spiral galaxies is  
12+log(O/H) $\sim$ 8.75. The central oxygen abundance in the Milky Way 
is similar to that in other large spirals.
Some fraction of the oxygen is locked into dust grains in H\,{\sc ii} regions.
\citet{estebanetal98} found that the fraction of the dust-phase oxygen 
abundance in the Orion nebula is about 0.08 dex. 
Then, the true gas+dust maximum value of the oxygen abundance in 
H\,{\sc ii} regions of spiral galaxies is 12+log(O/H) $\sim$ 8.85. 
This value can increase up to $\sim$ 8.90 -- 8.95 if  
temperature fluctuations inside H\,{\sc ii} regions are important. 
 
\subsection*{Acknowledgments}

We thank Yuri Izotov and Grazyna Stasi\'{n}ska for providing us with a new 
set of T$_{\rm e}$-equations in advance of publication. 
We thank the anonymous referee for helpful comments. 
The research described in this publication was made possible 
in part by Award No UP1-2551-KV-03 of the U.S. Civilian 
Research \& Development Foundation for the Independent States of the Former 
Soviet Union (CRDF). L.S.P was also partly supported by grant No 02.07/00132 
from the Ukrainian Fund of Fundamental Investigations.

\end{document}